\documentclass[lettersize,journal]{IEEEtran}
\usepackage{amsmath,amsfonts}
\usepackage{algorithmic}
\usepackage{algorithm}
\usepackage{array}
\usepackage[caption=false,font=normalsize,labelfont=sf,textfont=sf]{subfig}
\usepackage{textcomp}
\usepackage{stfloats}
\usepackage{url}
\usepackage{verbatim}
\usepackage{graphicx}
\usepackage{caption}
\usepackage{hyperref}

\begin{document}

\title{Data-Driven Linearization based Arc Fault Prediction in Medium Voltage Electrical Distribution System}

\author{{Mihir Sinha, Kriti Thakur, Prasanta K. Panigrahi, Alivelu Manga Parimi, Mayukha Pal}
 

\thanks{(Corresponding author: Mayukha Pal)}
\thanks{Mr. Mihir Sinha is a Data Science Research Intern at ABB Ability Innovation Center, Hyderabad 500084, India and also a B.Tech undergraduate student from Department of Mechanical Engineering, Indian Institute of Technology, Tirupati 517619, IN.}
\thanks{Ms. Kriti Thakur is a Data Science Research Intern at ABB Ability Innovation Center, Hyderabad 500084, India, and also a Ph.D. Research Scholar at the Department of Electrical and Electronics Engineering, BITS Pilani Hyderabad Campus, Hyderabad 500078, IN.}
\thanks{Prof. Prasanta K. Panigrahi is a Professor in the Department of Physical Sciences,Indian Institute of Science Education and Research, Kolkata, Mohanpur, Nadia,741246,IN.}
\thanks{Dr. Alivelu Manga Parimi is a Professor in the Department of Electrical and Electronics Engineering, BITS Pilani Hyderabad Campus, Hyderabad 500078, IN.}

\thanks{Dr. Mayukha Pal is with ABB Ability Innovation Center, Hyderabad-500084, IN, working as Global R\&D Leader – Cloud \& Advanced Analytics (e-mail: mayukha.pal@in.abb.com).}
}

\maketitle

\begin{abstract}
High-impedance arc faults (HIAFs) in medium-voltage electrical distribution systems are difficult to detect due to their low fault current levels and nonlinear transient behavior.  Traditional detection algorithms generally struggle with predictions under dynamic waveform scenarios.  This research provides our approach of using a unique data-driven linearization (DDL) framework for early prediction of HIAFs, giving both interpretability and scalability.  The proposed method translates nonlinear current waveforms into a linearized space using coordinate embeddings and polynomial transformation, enabling precise modelling of fault precursors.The total duration of the test waveform is 0.5 seconds, within which the arc fault occurs between 0.2 seconds to 0.3 seconds. Our proposed approach using DDL, trained solely on the pre-fault healthy region (0.10 seconds to 0.18 seconds) effectively captures certain invisible fault precursors, to accurately predict the onset of fault at 0.189 seconds, which is approximately 0.011 seconds (i.e., 11 milliseconds) earlier than the actual fault occurrence.  In particular, the framework predicts the start of arc faults at 0.189 seconds, significantly earlier of the actual fault incidence at 0.200 seconds, demonstrating substantial early warning capability. Performance evaluation comprises eigenvalue analysis, prediction error measures, error growth rate and waveform regeneration fidelity. Such early prediction proves that the model is capable of correctly foreseeing faults which is especially helpful in preventing real-world faults and accidents. It confirms that our proposed approach reliably predicts arc faults in medium-voltage power distribution systems.

\begin{figure}[H]
    \centering
    \captionsetup{justification=centering}
    \includegraphics[width=0.40\textwidth]{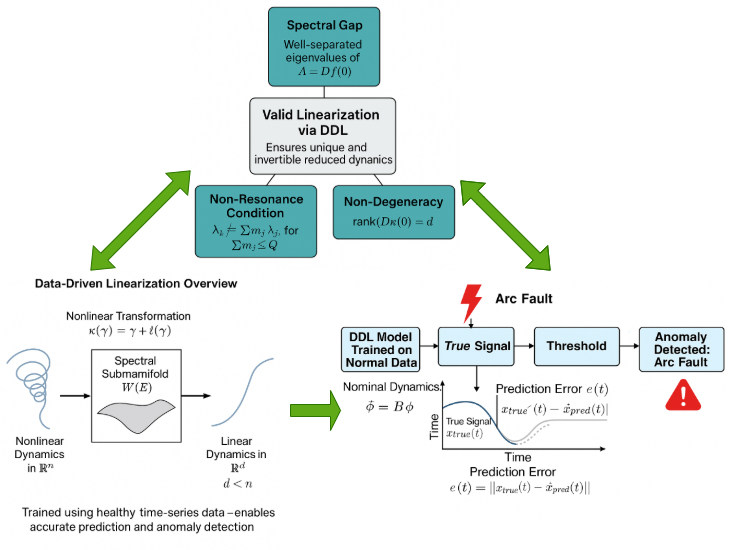}
    \caption{Conceptual overview of arc fault prediction using our DDL based approach}
    \label{fig:ddl_prediction}
\end{figure}

\end{abstract}

\section{Introduction}

\IEEEPARstart{E}lectricity plays an essential role in supporting modern society, powering homes, industrial sectors, and critical infrastructure.  The dependability and safety of power distribution systems are consequently of vital significance, especially for medium-voltage (MV) networks, which act as the backbone between high-voltage transmission and low-voltage consumer distribution. As electrical distribution networks become more complex, their need on reliable fault protection technologies grows proportionately.  Faults within medium-voltage (MV) networks can lead to serious consequences, such as damage to equipment, service disruptions, and possible dangers to human safety.  Conventional protection devices, such as overcurrent relays and fuses, function in a responsive way, acting only after a fault has occurred.  Therefore, there is a rising need to develop proactive solutions capable of recognising and reducing errors at an early stage. Protecting the electrical distribution system is therefore a crucial concern \cite{MP_USpatent, MP_WOpatent}.

In medium voltage distribution systems, arc faults, particularly high-impedance arc faults (HIAFs), pose significant challenges due to their small electrical fingerprints and erratic characteristics. According to UL Standard UL1699 \cite{UL1699}, an arc defect is characterised as a sustained release of electrical energy manifested as light through an insulating media, sometimes accompanied by the partial vaporisation of the electrodes.  This condition may arise from various contributing reasons, such as insulation deterioration in the electrical circuit, damaged electrical connections, increasing ambient humidity, or reduced dielectric strength of the conductor's insulation material.
Arc faults are broadly categorized into three major types \cite{thakur2023advancements}.  A series arc fault develops due to a discontinuity in a single conductor, usually resulting from a break in a power line.  A parallel arc fault occurs due to insulation breakdown between multiple conductors running in parallel, typically induced by thermal stress or external mechanical forces.  Ground arc faults arise when an unintentional electrical pathway is created between a current-carrying conductor and the ground, establishing an electrical connection with the earth's surface.
A high impedance fault (HIF) is a typical kind of single-phase grounding fault, generally induced by a wire encountering a high-resistance surface.  This commonly occurs in arcing that results in heat and light emissions, thus the term high-impedance arc fault (HIAF) \cite{zeng2023high}.  The increased integration of varied energy sources onto the grid has further complicated operational circumstances, hence enhancing the likelihood of HIAF events \cite{hyun2022arc}.

Over the years, several detection methods have been developed for detecting arc faults using traditional approaches like time domain-based \cite{9756255}, frequency-based\cite{dolkegowski2021novel}, time-frequency-based\cite{en16031256} arc fault detection such as Fast Fourier Transform (FFT), Discrete Wavelet Transform (DWT) and Empirical Mode Decomposition (EMD). The purpose is to identify frequency or time-frequency deviations in current and voltage signals following a fault. Traditional arc fault detection approaches suffer from poor generalization across variable loads, high sensitivity to noise, and limited real-time applicability. To enhance the robustness of fault detection systems, numerous researchers have investigated the application of machine learning (ML) and artificial intelligence (AI) techniques. Techniques such as support vector machines (SVM) \cite{dang2021series}, k-nearest neighbors (k-NN), and convolutional neural networks (CNN) \cite{s23177646} have demonstrated promising performance in effectively detecting arc faults, often yielding significantly higher accuracy compared to traditional methods. AI-based arc fault detection techniques, while promising higher accuracy, frequently require enormous amounts of labeled fault data and suffer from poor generalization across varied operating conditions.
However, most of the existing techniques are designed for post-fault detection, implying that faults are identified only after they have already happened.

This limitation emphasises the major motivation of this work to develop a prediction system capable of recognising arc faults before their actual appearance.  By transitioning from a reactive to a proactive protection strategy, the suggested technique intends to enhance system dependability and safety by enabling early intervention and lowering the risks associated with undetected incipient faults.

The work in \cite{MP_USpatentearc}  proposes a fundamental methodology for arc prediction, emphasising the significance of early detection measures in electrical systems. In this work, we offer a novel data-driven linearization (DDL) approach by \cite{haller2024data} for the early prediction of HIAF in MV distribution systems. 
We adopt this framework in this work due of its remarkable capacity to simplify the modeling of complex nonlinear systems.  First, DDL permits the transformation of a nonlinear system into a lower-dimensional linear representation by projecting it onto spectral submanifolds, designated as $\mathcal{W}(E)$.  These submanifolds are invariant, low-dimensional structures that successfully capture the prevailing dynamics of the system, allowing for a more interpretable and computationally efficient representation without considerable loss of important behavior.  This method not only simplifies the modeling process but also promotes generalizability across diverse operational conditions.

 Second, a fundamental feature of the DDL framework is in its forced response prediction capacity, which allows the model to properly estimate the system's future states even under external shocks.  This predictive resilience is crucial for early fault prediction, since it ensures that deviations from expected behavior, such as those induced by HIAF could be predicted immediately and accurately.  Together, these properties make DDL a strong and unique tool for predictive fault diagnosis in MV distribution systems, giving a combination of interpretability, scalability, and accuracy that is rarely obtained in traditional or black-box AI-based models.

The main contributions of this work are as follows:
\begin{itemize}
    \item The study introduces a unique data-driven linearization (DDL) framework for predicting HIAF in MV electrical distribution systems, employing only healthy current waveform data and eliminating the need for labeled fault instances.
    \item We demonstrate the use of prediction error and error growth rate as a robust and quantitative early warning indicator, displaying great effectiveness in predicting the initial beginning of HIAFs.
    \item We validate the proposed model on simulated MV current signals and show that it could anticipate HIAF up to 11 milliseconds before their actual occurrence, demonstrating superior performance over traditional reactive detection schemes.
\end{itemize}

The remainder of this paper is organized as follows: \\
Section~II reviews about the simulation setup of Arc Fault Modeling. Section~III presents the overview of DDL and its Application to Forced Response Prediction.
Section~IV describes our Methodology of DDL based Arc Fault Prediction. Section~V discusses about the Results and Discussions of Prediction Model. Finally, Section~VI concludes the study and outlines potential research directions.

\section{Simulation Setup}
\begin{figure*}[!htbp]
    \centering
    \includegraphics[width=17cm]{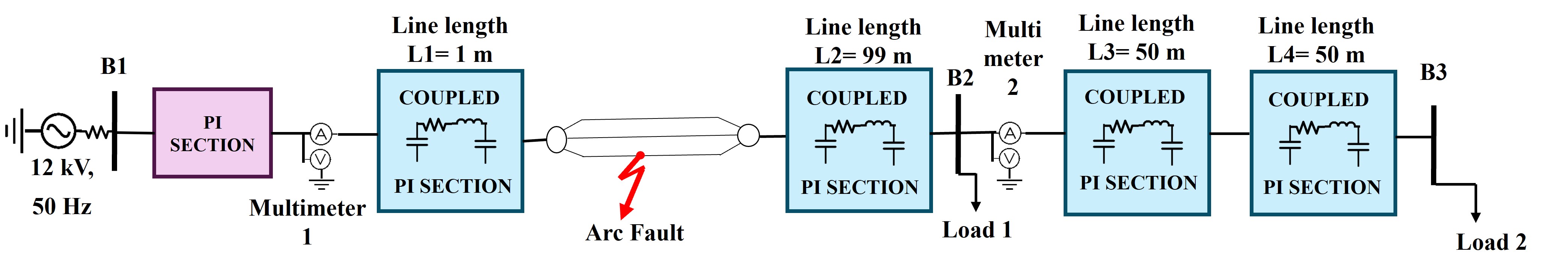}
    \caption{Simulation setup of a medium-voltage distribution network incorporating an arc fault.}
    \label{fig:system}
\end{figure*}

A medium-voltage (MV) distribution system has been developed and studied, running at a nominal voltage level of 12 kV and a frequency of 50 Hz.  The simulation environment utilised for this work is PSCAD, a widely utilised tool for power system transient analysis.  The system architecture comprises of three buses, designated as B1, B2, and B3, which act as critical nodes within the network.  Two loads, each rated at 10 MW, are connected to the system illustrated in Figure \ref{fig:system}, giving a realistic load situation for examining fault behavior.

To precisely replicate arc fault circumstances inside the distribution system, a high-impedance arc fault (HIAF) model has been implemented. The simulation framework integrates an arc fault model developed from the work of Wei et al \cite{wei2020distortion}, which serves as a basic reference for precisely modeling the dynamic behavior of arc faults in MV systems. This model has been updated and integrated into the simulation environment to recreate the transient and nonlinear electrical properties associated with high-impedance arc faults.  The adoption of this specific arc model ensures that the simulated arc events reflect the distortions and complexities frequently encountered in practical distribution networks.
To properly examine the effectiveness of the arc fault prediction system, arc faults were purposely produced at multiple sites throughout the MV distribution network.  Specifically, fault situations were simulated along the lines between Bus 1 and Bus 2, and between Bus 2 and Bus 3.  The deliberate placement of faults across different portions of the network enables the evaluation of the detection algorithm’s sensitivity, generalizability, and spatial accuracy.  By altering the position of the faults, the simulation captures a wide collection of fault-induced current signatures, which are critical for training and evaluating the model under different electrical situations.
To enable high-resolution time-domain study of arc phenomena, the simulations were conducted with a high sampling rate of 20 kHz, which corresponds to a temporal resolution of 50 microseconds per sample.  This  temporal resolution allows the collection of rapid changes in current and voltage waveforms induced by arc ignition and extinction processes, which are crucial for effective feature extraction and fault prediction.
All algorithmic evaluations, including signal preprocessing, feature extraction, and machine learning model inference, were performed in Python.  The studies were conducted on a conventional computing setup comprising an Intel Core i5 CPU running at 2.60 GHz and 16 GB of RAM.
Each simulation scenario is run over a total time of 0.5 seconds to capture both pre-fault and post-fault system behavior completely. An arc fault is injected into the system at 0.2 seconds from the start of the simulation and is sustained for a duration of 0.1 seconds, effectively deactivating at 0.3 seconds. This timing guarantees that the simulation includes a significant window of steady-state operation before the fault occurs, allowing for a clear observation of the transient reaction during and after the fault event. The goal is to verify that the created model successfully predict faults across a wide range of realistic circumstances experienced in medium-voltage distribution systems.

\section{Overview of DDL and Its Application to Forced Response Prediction}

\subsection{Overview of DDL Framework}

Using DDL, one could replace a nonlinear system with a simpler, linear one that exists on spectra submanifolds, denoted $\mathcal{W}(E)$ which are low-dimensional, remain invariant and collect the major system behaviors. Using a nonlinear transformation on the data gives DDL the skill to predict accurately and detect early anomalies, even when other influences are difficult to see.\cite{haller2024data}

\subsection{DDL for Autonomous Systems}

Consider an autonomous nonlinear system\cite{haller2024data} described by
\begin{equation}
\dot{y}(t) = f(y(t)), \quad y(t) \in \mathbb{R}^n
\label{eq:autonomous}
\end{equation}
where \( y(t) \) is the state vector at time \( t \), and \( f: \mathbb{R}^n \rightarrow \mathbb{R}^n \) is a smooth, nonlinear vector field. The goal of DDL is to find a transformation into reduced coordinates \( \gamma \in \mathbb{R}^d \), where \( d \ll n \), such that the transformed state \( \varphi \in \mathbb{R}^d \) on $\mathcal{W}(E)$, evolves approximately linearly. This transformation is defined as
\begin{equation}
\varphi = \kappa(\gamma) = \gamma + \ell(\gamma)
\label{eq:kappa}
\end{equation}
where \( \ell: \mathbb{R}^d \rightarrow \mathbb{R}^d \) is a nonlinear correction function, typically a multivariate polynomial. In the new coordinates, the dynamics are approximated as
\begin{equation}
\dot{\varphi} = B\varphi
\label{eq:linear}
\end{equation}
with \( B \in \mathbb{R}^{d \times d} \) being the reduced-order linear operator. If we want to conserve these dynamics, the transformation must obey the invariance PDE:
\begin{equation}
D_\gamma \ell(\gamma) \cdot B\gamma + f(\gamma + \ell(\gamma)) = B(\gamma + \ell(\gamma))
\label{eq:invariance}
\end{equation}
where \( D_\gamma \ell(\gamma) \in \mathbb{R}^{d \times d} \) is the Jacobian of \( \ell(\gamma) \).

\subsection{Conditions for Valid Linearization}

To guarantee the existence and uniqueness of the transformation \( \kappa \) and the linear operator \( B \), the following mathematical conditions must hold near equilibria\cite{haller2024data}. First, a spectral gap must exist in the Jacobian matrix \( A = Df(0) \), which separates slow (dominant) modes from fast transients. Second, a \textit{non-resonance condition} is required:
\begin{equation}
\lambda_k \neq \sum_{j=1}^{n} m_j \lambda_j, \quad \text{for all } \sum m_j \leq Q
\label{eq:nonresonance}
\end{equation}
where \( \lambda_j \) are eigenvalues of \( A \), \( m_j \in \mathbb{N} \), and \( Q \in \mathbb{N} \) is the maximum polynomial degree. Third, a \textit{non-degeneracy condition} ensures that \( \kappa \) is invertible near the equilibrium:
\begin{equation}
\text{rank}(D\kappa(0)) = d
\label{eq:rank}
\end{equation}
Under these assumptions, the nonlinear dynamics in transformed coordinates could be written as
\begin{equation}
\dot{\varphi} = B\varphi + q(\varphi)
\label{eq:nonlinear_correction}
\end{equation}
where \( q(\varphi) \) contains higher-order nonlinear corrections.

\subsection*{D. DDL vs. Extended Dynamic Mode Decomposition (EDMD)}

Both DDL and EDMD are aimed at model nonlinear behaviors as linear ones although each of them uses different concepts and approaches—with main differences in prediction and understanding tasks.

A large number of high-dimensional characteristics is created in EDMD \cite{Ouala_2023} by using well-known basis functions such as polynomials or Fourier terms, on the data. The situation in the lifted space is then analyzed using linear techniques. Nevertheless, one important difficulty with EDMD is that mapping data to the feature space is usually not reversible. Even though EDMD gives an approximate linear relationship in feature space, it is not always possible to return the features to the original, raw signal space. Since not all states are recoverable, EDMD is less helpful for tasks where we must reconstruct the initial signal.
\cite{haller2024data}

This means DDL is designed specifically to be able to transition back to the original data. The transition from the reduced space to observed variables is found as a polynomial function and its inverse is given in terms of a formal power series. It means that to transform from latent coordinates $\varphi$ to reduced coordinates $\gamma$, following formula needs to be used:

\begin{equation}
\gamma = \kappa^{-1}(\varphi) = \varphi + \sum_{|k|=2}^{\infty} q_k \varphi^k
\end{equation}

By using this approach, the transformation takes place easily near the origin and is reversible there. Substituting this expression into the linear dynamics $\dot{\gamma} = B\gamma$, we obtain the dynamics in $\varphi$-coordinates as:

\begin{equation}
\dot{\varphi} = B\varphi + \sum_{|k|=2}^{\infty} Bq_k \varphi^k
\end{equation}

This result shows that the nonlinear behavior of the system is captured in a structured manner through the higher-order terms in $\varphi$, while the leading-order term remains linear. To make the model tractable for computation and simulation, the infinite series is truncated at a certain polynomial degree $r$. This results in a stacked system representation:

\begin{equation}
\frac{d}{dt} 
\begin{bmatrix}
\varphi \\
K_{\geq 2}(\varphi)
\end{bmatrix}
=
\begin{bmatrix}
B & BQ_2 \\
0 & 0
\end{bmatrix}
\begin{bmatrix}
\varphi \\
K_{\geq 2}(\varphi)
\end{bmatrix}
\end{equation}

Here, $K_{\geq 2}(\varphi) \in \mathbb{R}^r$ represents the vector of monomials of degree 2 or higher (e.g., $\varphi_1^2$, $\varphi_1\varphi_2$, $\varphi_2^2$, etc.), and $Q_2 \in \mathbb{R}^{d \times r}$ maps their influence on the evolution of the main latent dynamics.

With this approach, DDL delivers an edge over EDMD: it allows information to move in both directions between the latent and signal layers. As a result, it becomes possible to predict system behavior in the original coordinates which is important for tools that predict early faults. Besides, the behavior of the hidden layers is easy to explain and gives important physical information about the system.

\subsection*{D. Forced Response Prediction via DDL}

The DDL framework could still accurately predict how the system will respond to disturbances\cite{haller2024data}, thanks to its training only on healthy data. That’s why DDL helps greatly when trying to simulate situations such as arc faults which commonly arise after ordinary operation has started

Consider a nonlinear, non-autonomous system with a small time-dependent external input:

\begin{equation}
\dot{x}(t) = f(x(t)) + \epsilon F(x(t), t), \quad \epsilon \ll 1
\end{equation}

In this case, \( f(x(t)) \) lists the dynamics when there is no disturbance, \( F(x(t), t) \) represents surprises such as arc faults and \( \epsilon \) measures the amount of disturbance which is considered small.

Applying the nonlinear transformation \( \varphi = \kappa(\gamma) = \gamma + \ell(\gamma) \)\cite{haller2024data}, the dynamics in the latent space evolve as:

\begin{equation}
\dot{\varphi}(t) = B\varphi(t) + q(\varphi(t)) + \epsilon \hat{F}(\varphi(t), t)
\end{equation}

the reduced linear dynamics are controlled by \( B\varphi(t) \), the next-order corrections by \( q(\varphi(t)) \) and \( \hat{F}(\varphi(t), t) \) is the external input expressed in latent space.

Transforming to reduced coordinates \( \gamma \), the dynamics become:

\begin{equation}
\dot{\gamma}(t) = B\gamma(t) + \epsilon (I + D\ell(\gamma(t)))^{-1} \hat{F}(0, t)
\end{equation}

With this setting, \( D\ell(\gamma) \) is the Jacobian of \( \ell(\gamma) \) and its inverse changes how much the forcing factors influence the system. The local process \( \hat{F}(0, t) \) considers the nearby and important factors affecting our system so it could react to external inputs.

\section{Methodology of DDL-Based Arc Fault Prediction}
\begin{figure*}[t]  
  \centering
  \includegraphics[width=1\textwidth]{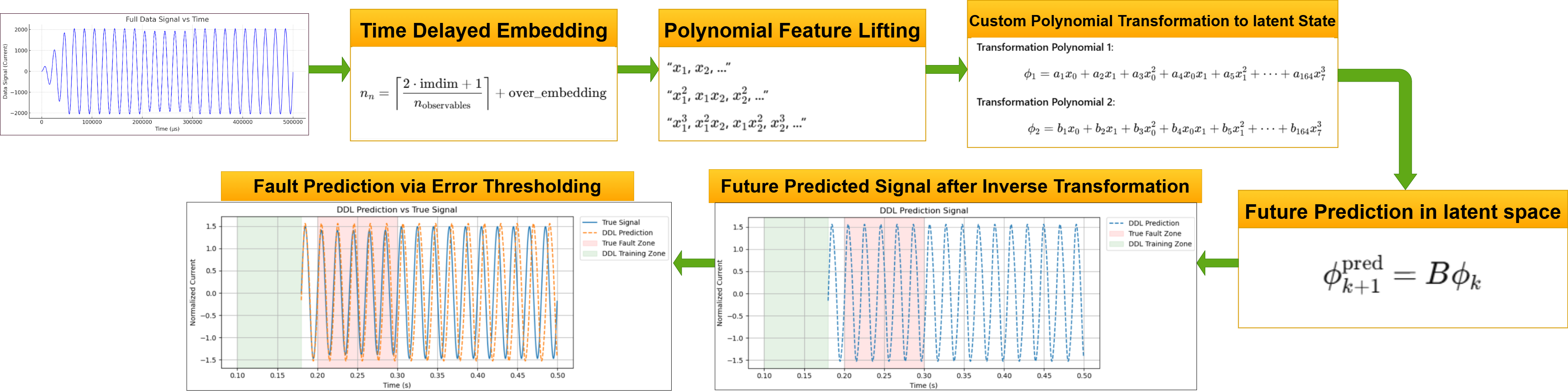} 
  \caption{DDL Flow Chart for Arc Fault Prediction}
  \label{fig:ddl_architecture}
\end{figure*}

The procedural flow for the proposed arc fault prediction framework is presented in Figure \ref{fig:ddl_architecture}.  This flowchart explains the sequential stages involved in executing the prediction model, beginning with the collecting of current waveform data under typical operating conditions.  The data is then preprocessed and turned into a high-dimensional polynomial feature space using the DDL approach.  A reduced-order linear model is subsequently created to capture the system’s prevailing dynamics.  During operation, the model continuously anticipates the expected system behavior, and variations between the projected and actual responses are monitored.  If the deviation exceeds a threshold, it is taken as an early warning indicator of a probable arc fault.  This structured technique offers effective and fast fault prediction, hence boosting the safety and dependability of medium-voltage distribution systems.
\subsection*{Step 1: Extract Healthy Data}

The initialization of our DDL-based framework uses a healthy current signal portion from the simulated dataset which was sampled at 50 micro-seconds intervals. The initial time span up to 0.1 seconds represents transience while the known defect emerges between 0.2 seconds to 0.3 seconds. The model training data selects the period from 0.1 to 0.18 seconds (2000 to 3600 samples) since it represents a faultless section.

Experts position this segment within the slow spectral submanifold $\mathcal{W}(E) \subset \mathbb{R}^n$ because it reveals main dynamics while neglecting transients. The training of this specific area allows the DDL model to discover an effective nonlinear transformation between the input x and the output $\phi = \kappa(x)$, where $\phi : \mathbb{R}^n \rightarrow \mathbb{R}^d$, with $d < n$.

\subsection*{Step 2: Time-Delay Embedding}

We make use of DDL-style time-delay embedding\cite{haller2024data} to capture how the input signal changes in an increased number of dimensions. We use embedding dimension of imdim = 2 and an over-embedding value of 3, to calculate the number of delay coordinates as:

\begin{equation}
n_n = \left\lceil \frac{2 \cdot \text{imdim} + 1}{n_{\text{observables}}} \right\rceil + \text{over\_embedding} = \left\lceil \frac{5}{1} \right\rceil + 3 = 8
\end{equation}

This results in each point of the signal being represented as a 8-dimensional embedded vector:

\begin{equation}
y_k = [x_k, x_{k+1}, x_{k+2}, \ldots, x_{k+7}]^T \in \mathbb{R}^8
\end{equation}

Such delay embedding preserves the topological structure of the original system's attractor, as guaranteed by Takens' embedding theorem. These embedded vectors form the input for the subsequent polynomial feature lifting step in the DDL framework.

\subsection*{Step 3: Polynomial Feature Lifting}
Each embedded vector is lifted using polynomial expansion of degree 2:
\begin{equation}
\
\Phi(y_k) \in \mathbb{R}^9
\
\label{eq:nominal}
\end{equation}
This produces 9 monomials, such as $x_k^2$, $x_k x_{k+1}$, $x_{k+1}^2$, forming the lifted feature matrix $K_k(\phi)$. This nonlinear mapping allows linear modeling in the lifted space.

\subsection*{Step 4: Latent Transformation Learning and Error Evaluation}

Here, the model explores and understands the evolution of the system using fewer, more simplified, representation of features. The matrix \( Q \in \mathbb{R}^{d \times (N - d)} \) brings the lifted polynomial features to a space of coordinates \( \varphi_k = QK_k \), where $\varphi = \phi \big|_{\mathcal{W}(E)} \in \mathbb{R}^d$ and $B \in \mathbb{R}^{d \times d}$ controls how they move linearly.\cite{haller2024data}

To ensure dynamic consistency, DDL minimizes the residual of the invariance equation, which aligns the predicted latent state evolution with the actual one:

\begin{equation}
\left[ I \; Q \right]
\begin{bmatrix}
\hat{\varphi}_k \\
\hat{K}_k
\end{bmatrix}
\approx
\left[ B \; BQ \right]
\begin{bmatrix}
\varphi_k \\
K_k
\end{bmatrix}
\label{eq:invariance}
\end{equation}

The model minimizes the prediction error, defined as:

\begin{equation}
\mathcal{L}(Q, B) = \left\| 
\left[ I \; Q \right] 
\begin{bmatrix}
\hat{\varphi}_k \\
\hat{K}_k
\end{bmatrix}
-
\left[ B \; BQ \right]
\begin{bmatrix}
\varphi_k \\
K_k
\end{bmatrix} 
\right\|^2
\label{eq:loss}
\end{equation}

A small residual shows that the linear model has fitted the dynamics of the system well. Meanwhile, huge mistakes may be caused by behavior that has not been modeled or faults that are beginning to appear. Also, since the eigenvalues of matrix \( B \) are usually close to 1, the dynamics in the latent space do not exhibit any long-term oscillations.

\subsection*{Step 5: Predict Future State}
Since the DDL model is trained , it predicts the system using the learned $B$ as:

\begin{equation}
\
\varphi^{\text{pred}}_{k+1} = B \varphi_k
\
\label{eq:nominal}
\end{equation}
Starting from $\varphi_0$, this simulates the forward evolution of the healthy signal.

\subsection*{Step 6: Inverse Transformation}
The results from the latent state prediction enter the observable range through reverse mapping:
\begin{equation}
\
x_k^{\text{pred}} = \varphi_k + Q_{\text{inv}} K_k
\
\label{eq:nominal}
\end{equation}
Signal reconstruction takes place here to make possible comparisons with original measurements.

\subsection*{Step 7: Fault Detection via Error Thresholding}
A fault is flagged when the absolute error or its derivative exceeds a threshold:
\begin{equation}
\
|x_k - \hat{x}_k| > \theta, \quad \text{or} \quad \frac{d}{dt}|x_k - \hat{x}_k| > \delta
\
\label{eq:nominal}
\end{equation}
Here in our case, thresholds $\theta$ and $\delta$ exist either from empirical studies or statistical determination (such as $\mu + 3\sigma$). The system detects faults at an early stage prior to observable waveform changes through this mechanism.

\section{Results and Discussion}

\begin{figure}[H]
    \centering
    \captionsetup{justification=centering}
    \includegraphics[width=0.48\textwidth]{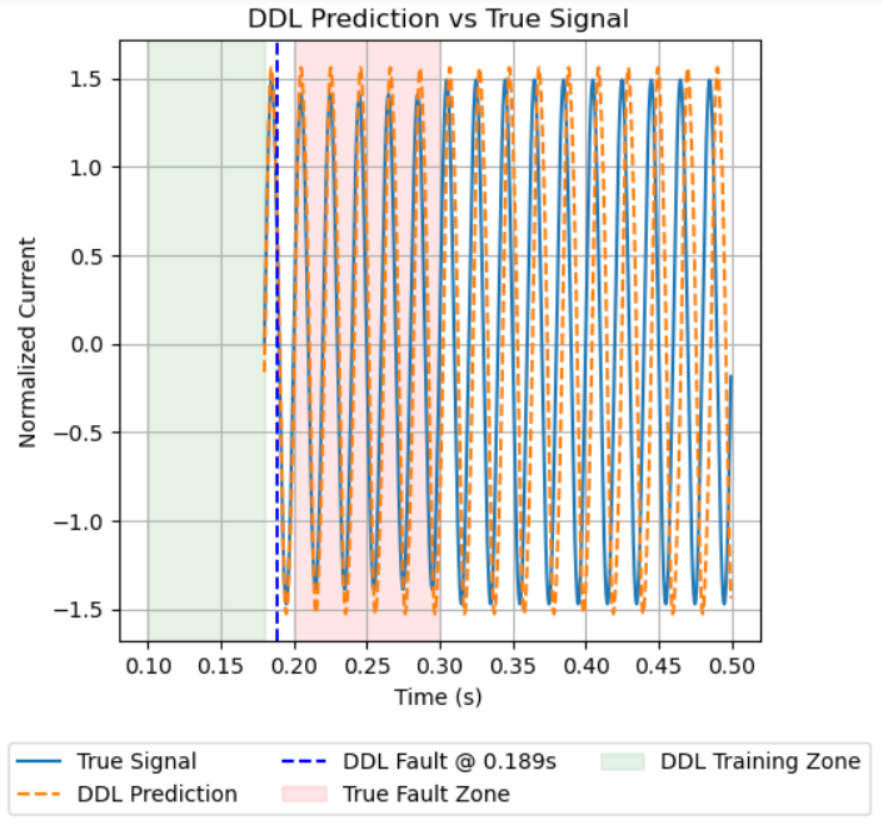}
    \caption{DDL Prediction vs True Signal.}
    \label{fig:ddl_prediction}
\end{figure}

\subsection{Predicted vs Actual Behavior}
Model performance evaluation happens through comparison between computed and actual current waveforms in healthy and faulty conditions. Long-term operation of the DDL model starts with training it using faultless data between 0.10 – 0.18 seconds then applies its learned latent features for extrapolation. Figure \ref{fig:ddl_prediction} depicts the healthy training area (green) together with the predicted signal (orange dashed line) maintaining close proximity to the actual signal (blue). The divergence starts at 0.189 seconds before the known fault initiation point at 0.200 seconds producing a warning that starts 11 milliseconds early. The divergence of absolute error thresholds along with its derivative signals this predictive behavior of DDL.

\subsection{Prediction Error Evolution}

When prediction error shows changes it reveals sensitive faults. Predicted signals are compared with true signals inside the observable space to determine residuals. The model indicates rising prediction errors whenever system faults occur because its training involves healthy data only.

\begin{figure}[!htbp]
\centering
\subfloat[]{\includegraphics[width=2.5in]{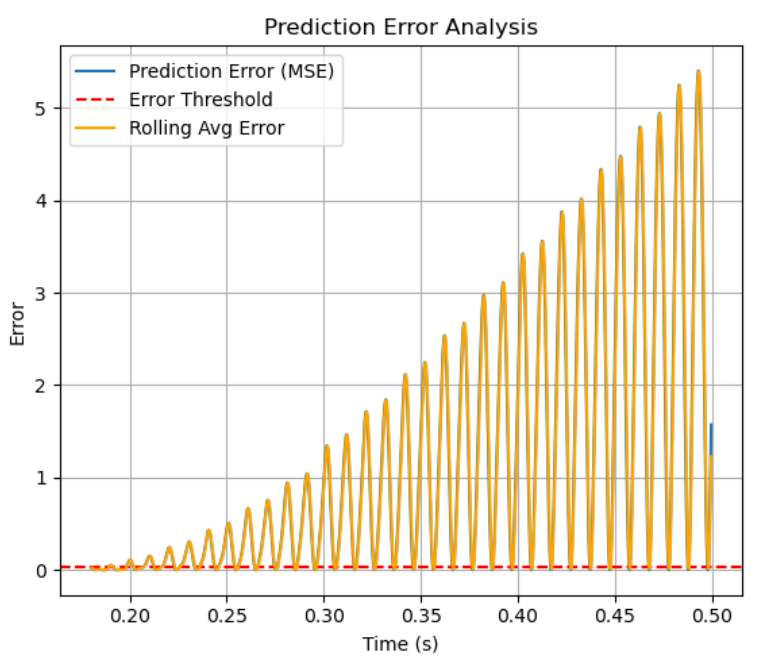}%
\label{fig_first_case}}
\hfil
\subfloat[]{\includegraphics[width=2.5in]{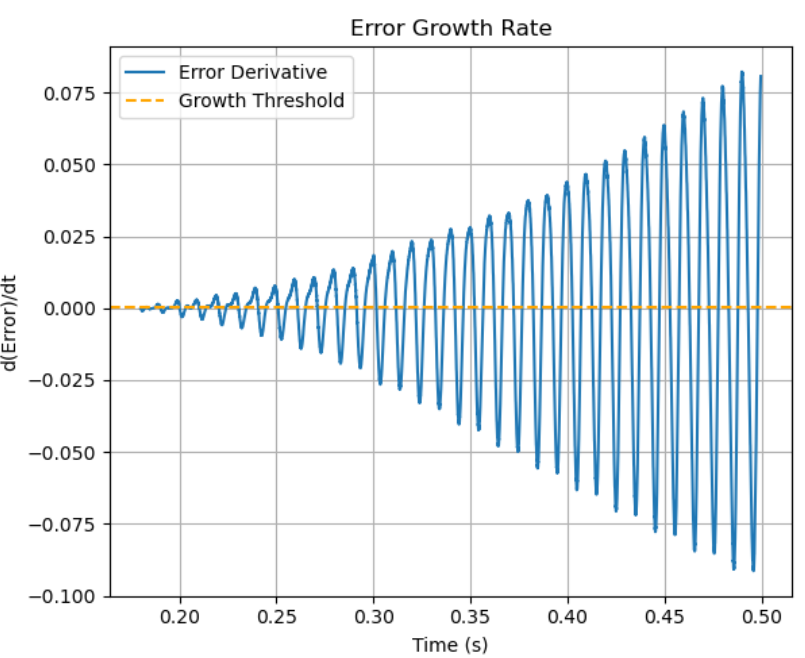}}%
\label{fig_second_case}
\caption{Prediction Error (MSE) and its Growth Rate. Red dashed lines represent thresholds for fault detection. (a) Prediction Error. (b) Error Growth Rate.}
\label{fig_sim}
\end{figure}

The detection of process deviations uses a $3\sigma$ control limit which sets significant limits with sample number one through thirty to prevent unnecessary alerts. The system monitors error-derived average and error derivative values through a rolling algorithm in order to detect faults based on set thresholds. Figure \ref{fig_sim} depicts this mechanism. From 11 milliseconds before the fault develops the system began showing separation at 0.189 seconds.
\begin{figure}[H]
    \centering
    \captionsetup{justification=centering}
    \includegraphics[width=0.48\textwidth]{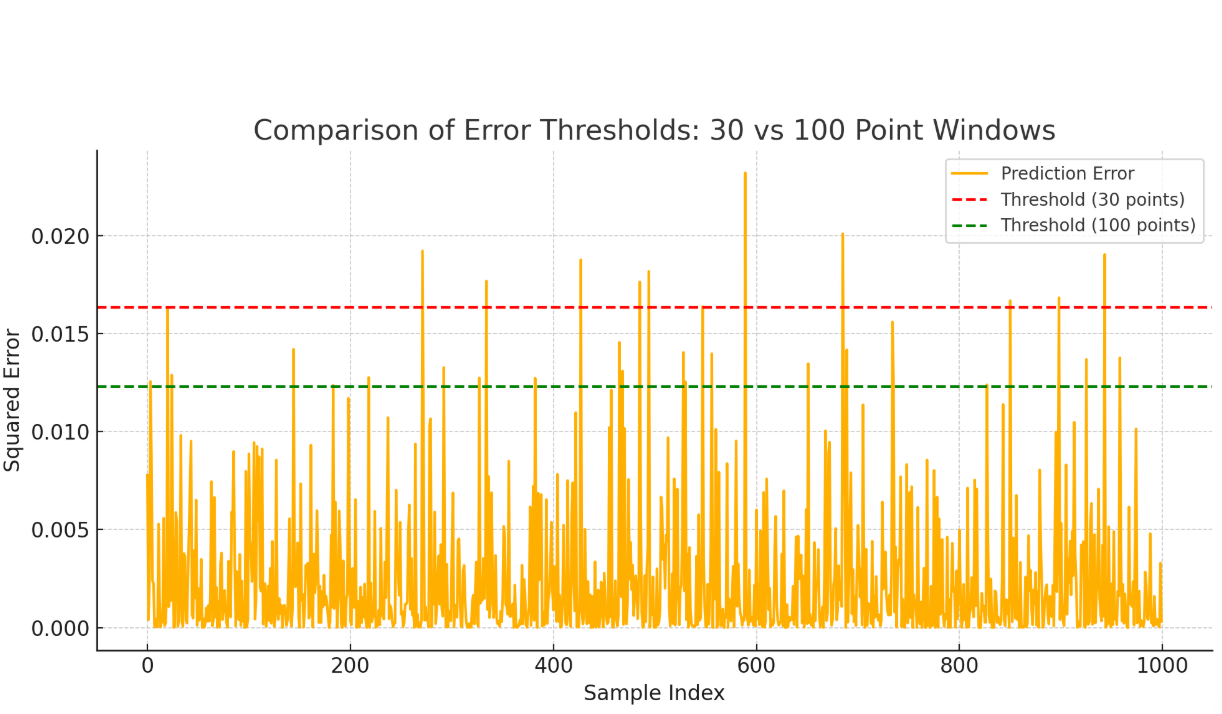}
    \caption{Error Threshold Comparison}
    \label{fig:ddl_pred}
\end{figure}

As shown in Figure \ref{fig:ddl_pred}, the size of the moving window used to compute the error threshold ($\mu + 3\sigma$) significantly influences fault detection behavior. The quick response of a 30-point window to sudden changes in prediction error means faults could be identified more quickly. On the other hand such high sensitivity means that the device could pick up temporary noise in the atmosphere which may mistake it for a valid signal. Meanwhile, a bigger 100-point window leads to less impacts from quick changes in the signal, but there is a delay in identifying a fault.

In our implementation, we adopt a 30-point window to prioritize early fault prediction, which is critical in time-sensitive electrical fault scenarios. To mitigate noise-induced false alarms despite the short window, we incorporate a rolling average filter on the prediction error  and apply an additional threshold on its derivative, as shown in Figure \ref{fig_sim}. This two-layer strategy ensures that only sustained and growing deviations from healthy dynamics are flagged, thereby maintaining high sensitivity without compromising reliability.

\subsection{Polynomial Lifting and Nonlinear Transformation}
Using time-delay embedding ($d=2$, over-embedding=5), signals are transformed into 8D vectors. These are lifted via third-degree polynomial expansion into a 164-dimensional feature space, including monomials like $x_0^2$, $x_1x_3$, and $x_2^3$.

The lifted features are linearly combined through a learned matrix $H \in \mathbb{R}^{2 \times 164}$ to obtain latent variables:
\begin{equation}
\
\phi_1 = H_1 \cdot \Phi(x), \quad \phi_2 = H_2 \cdot \Phi(x)
\
\label{eq:nominal}
\end{equation}
The custom adaptable transformation mechanism fits the system dynamics to optimize fault detection accuracy.

\subsection{Latent Linear Dynamics and Eigenvalue Analysis}
The latent dynamics evolve linearly as\cite{haller2024data}:
\begin{equation}
\
\varphi_{k+1} = B \varphi_k
\
\label{eq:nominal}
\end{equation}
with $B \in \mathbb{R}^{2 \times 2}$. The eigenvalues $\lambda_{1,2} = 1.044 \pm 0.079j$ indicate mild instability and oscillations.

Under the analysis of 50 selected modes from the full lifted system of 164-dimension, the spectral decomposition reveals that 43 modes are unstable ($|\lambda| > 1$), 6 modes are decaying ($|\lambda| < 1$), and only 1 mode is neutral ($|\lambda| = 1$). This distribution highlights the presence of significant unstable and decaying modes in the lifted feature space, which must be stabilize to achieve stable long-term prediction.

To stabilize the latent evolution, the eigenvalues of the learned dynamics matrix $B \in \mathbb{R}^{2 \times 2}$ are projected onto the unit circle:
\begin{equation}
\
\lambda_{\text{confined}} = e^{i\theta}, \quad B_{\text{osc}} = V \, \text{diag}(e^{i\theta}) \, V^{-1}
\
\label{eq:nominal}
\end{equation}

Here, V is still the original eigen vectors matrix of $B$, of the learned dynamics. This transformation preserves the system’s oscillatory frequency while eliminating amplitude drift, allowing the model to generate bounded, sustained oscillations. Consequently, any deviation from this stable trajectory during testing is attributable to true external perturbations, such as arc faults, enabling reliable early prediction.

\subsection{Early Divergence Prior to Fault Onset}

In our case study, although the arc fault officially begins at $t = 0.200$ seconds, the prediction error starts rising significantly at $t = 0.189$ seconds—an early warning triggered by subtle, non-visible system perturbations. These include minor phase shifts, impedance fluctuations, or waveform damping that precede the actual arc event. Importantly, such precursors are not visually evident in the signal but become detectable in the latent space where DDL performs its modeling.

This early divergence is a direct consequence of the DDL construction. By learning a reduced-order linear approximation from purely healthy data with some latent forcing conditions which are invisible in the waveform, DDL builds a model of the system’s expected future trajectory. Any deviation from this learned behavior, however small, is captured by rising prediction errors and their derivatives. As thresholds are determined using a short post-training segment (e.g., 30 samples), even minute pre-fault forcing conditions embedded within the training window could trigger early anomaly detection , when the error and error growth rate starts crossing their respected thresholds.

This phenomenon was validated by comparing two scenarios: one in which a fault was present between $0.2$–$0.3$ seconds, and another in which that same segment was removed to simulate a fully healthy waveform. The fault-present case exhibited a sharper and earlier error rise with satisfactory error growth rate as compared to the fault-free case where it happened later at 0.223 s with unsatisfactory rate of change of error, confirming that DDL is capable of identifying incipient disturbances before they result in visible waveform distortion.

On the basis of the above comparison of the two scenes, it can be concluded that setting the thresholds for error and error growth rate are some of major factors involved in making the effective prediction of arc fault before its occurrence. 

\subsection{Effect of Healthy Training Duration}

The impact of healthy training duration on fault prediction is summarized in Table~\ref{tab:training_duration_effect}.

\begin{table}[H]
\centering
\caption{Impact of Training Window Duration on Predicted Fault Timing}
\label{tab:training_duration_effect}
\begin{tabular}{|c|c|c|}
\hline
\textbf{Training End (s)} & \textbf{Healthy Duration (s)} & \textbf{Predicted Fault (s)} \\ \hline
0.110 & 0.010 & 0.112 \\ \hline
0.120 & 0.020 & 0.123 \\ \hline
0.130 & 0.030 & 0.183 \\ \hline
0.140 & 0.040 & 0.202 \\ \hline
0.150 & 0.050 & 0.202 \\ \hline
0.160 & 0.060 & 0.202 \\ \hline
0.170 & 0.070 & 0.178 \\ \hline
0.180 & 0.080 & 0.189 \\ \hline
0.190 & 0.090 & 0.203 \\ \hline
\end{tabular}
\end{table}

From this analysis, four distinct regimes emerge:

\begin{itemize}
    \item Underfitting Region (0.11--0.13 s): With a brief training window, the system lacks plenty of data on healthy behavior which may cause the predictions to be unstable or easily triggered by false positive scenarios.
    
    \item Delayed detection (0.14--0.16 s): The model captures the healthy dynamics reasonably well, but because the training ends very early to the fault onset, it lacks a safety buffer to distinguish between normal and prefault signals. This results in a fault prediction that occurs nearly at or after the actual fault start time.
    
    \item Optimal Detection Window (0.17--0.18 s): This region strikes a balance between capturing sufficient healthy behavior and maintaining a clean separation from prefault conditions, but also capturing sufficient invisible forcing effects. The model trained up to 0.18 s predicts the fault at 0.189 s, offering a reliable early warning approximately 11 milliseconds ahead of fault onset. This supports the claim that DDL anticipates faults by detecting latent divergences before they become visible in the waveform.
    
    \item Overfitting Region (0.19 s): On Extending training window too close to or into the pre-fault zone introduces subtle forcing effects (e.g., early arc activity or impedance changes) into the model which leads to overfitting. This reduces sensitivity to future deviations of the errors from the threshold and delays fault detection, as the model begins treating early fault indicators as part of the normal behavior.
\end{itemize}

These results highlight that a training endpoint at 0.18 seconds provides the most effective compromise. It captures the system’s healthy dynamics while avoiding contamination from incipient fault signals. This leads to early, stable, and accurate fault prediction, reinforcing DDL’s strength as a proactive and robust fault prediction method.

\section{Conclusion}

This study shows DDL provides a reliable means to detect early arc faults in MV power distribution systems. While other systems react once waveform issues appear, DDL tries to foresee flaws by calculating a simpler model of system behavior. It watches for shifts in latent space, making it possible to detect problems much earlier than they appear in the original signal.

A key highlight of our results is DDL's ability to issue early warnings, even 11 milliseconds prior to actual fault onset. In our case study, although the arc fault officially started at 0.200 seconds, the prediction error began to rise sharply at 0.189 seconds. This early divergence is attributed to subtle pre-fault perturbations such as impedance variations, phase shifts, and waveform damping—phenomena that are not visually observable but become amplified in the latent space constructed by DDL.

We tested our prediction by looking at the results from a signal with a fault in the 0.2–0.3 s interval and then with that same segment removed from the signal, to demonstrate how it could be detected as healthy. The error threshold was crossed first in the presence of fault, at 0.189 s with sharp error growth rate, but in the fault-free case, it happened later at 0.223 s with unsatsfactory error growth rate. As a result, we could see that DDL reacts to issues that happen before a fault could be identified by any sensor. These results prove that DDL is capable of telling apart incipient abnormalities from normal behavior, helping to give early warning when arc fault is about to occur.

Time-delay embedding and third-degree polynomial feature lifting are applied by DDL for modeling to project one-dimensional signals into a 164-dimensional space. Features extracted in this high dimensional space are mapped by a learned transformation matrix to a 2D latent space in which linear growth is required. On the observation of 50 modes, we got to know that a small number of these drive the system’s overall development. Truncating these unsteady modes by eigenvalue confinement aids in obtaining proper and bounded predictions. As a result, DDL could forecast consistently over time and remove fleeting issues caused by noise.

The thresholding strategy was based on a 30-sample window immediately following the training zone. Though short, this window—combined with a rolling average filter and derivative thresholding—proved highly effective. It enabled the model to distinguish genuine divergence from healthy dynamics while remaining resilient to spurious fluctuations. Both the magnitude and growth rate of the prediction error served as highly sensitive indicators of emerging faults.

Our analysis of training window durations further reinforced the importance of training data selection. Training too early (before 0.13~s) led to underfitting and false alarms, while extending too close to the fault onset (e.g., to 0.19~s) caused overfitting and delayed detection. The optimal training endpoint was found to be 0.18~s, which offered the best separation between healthy dynamics and incipient pre-fault activity. Models trained with this configuration were able to detect upcoming faults early while remaining stable in fully healthy conditions.

Overall, DDL is not limited to detection; it actually anticipates system failures. The approach leverages five methods ,i.e., nonlinear polynomial lifting, latent space modeling, spectral filtering, proper training window selection and error thresholding to make effective fault prediction possible in real time using our DDL integrated approach. Thanks to its ability to respond to disturbances and stay strong in normal use, DDL is very useful in predictive maintenance for industry power systems. Future research might investigate to train this model on various real world high impedence arc fault data to get the optimized threshold value for error and error growth rate and selection of appropriate training window to make this prediction model most effective to predict arc faults before its occurence, in real time.


\bibliography{main}

\vfill

\end{document}